\documentclass[preprint,aps,prd,a4paper,tightenlines,nofootinbib,showpacs]{revtex4-2}
\usepackage{amsmath,amssymb,graphicx,mathrsfs,color}
\usepackage{bm} 
\usepackage{ulem}
\usepackage{feynmp}
\usepackage{ifpdf}
\ifpdf
\fi
\makeatletter
\def\endfmffile{%
  \fmfcmd{\p@rcent\space the end.^^J%
          end.^^J%
          endinput;}%
  \if@fmfio
    \immediate\closeout\@outfmf
  \fi
  \IfFileExists{\thefmffile.mp}{\immediate\write18{mpost \thefmffile}}{}
  \let\thefmffile\relax
}
\makeatother
\newcommand{\nn}{\nonumber\\}

\newcommand{\ben}{\begin{displaymath}}
\newcommand{\een}{\end{displaymath}}
\newcommand{\be}{\begin{equation}}
\newcommand{\ee}{\end{equation}}
\newcommand{\bea}{\begin{eqnarray}}
\newcommand{\eea}{\end{eqnarray}}

\newcommand{\A}{\alpha}
\newcommand{\B}{\beta}

\newcommand{\Gf}{G^{(4)}}
\newcommand{\Kf}{K^{(4)}}
\newcommand{\Gt}{G^{(2)}}

\newcommand{\q}{\bar{q}}
\newcommand{\D}{\bar{D}}

\newcommand{\bc}{\begin{center}}
\newcommand{\ec}{\end{center}}

\newcommand{\eqn}[1]{\label{#1}}
\newcommand{\eq}[1]{Eq.~(\ref{#1})}
\newcommand{\eqs}[1]{Eqs.~(\ref{#1})}
\newcommand{\fign}[1]{\label{#1}}
\newcommand{\fig}[1]{Fig.~\ref{#1}}

\begin{document}
\title{Covariant tetraquark equations in quantum field theory}
\author{A. N. Kvinikhidze}
\affiliation{Andrea Razmadze Mathematical Institute of Tbilisi State University, 6, Tamarashvili Str., 0186 Tbilisi, Georgia}
\email{sasha\_kvinikhidze@hotmail.com}
\author{B. Blankleider}
\affiliation{
College of Science and Engineering,
 Flinders University, Bedford Park, SA 5042, Australia}
\email{boris.blankleider@flinders.edu.au}

\date{\today}

\begin{abstract}

We derive general covariant coupled equations of QCD describing the tetraquark in terms of a mix of four-quark states $2q2\q$, and two-quark states $q\q$. The coupling of $2q2\q$ to $q\q$ states is achieved by a simple contraction of a four-quark $q\q$-irreducible Green function down to a two-quark $q\q$ Bethe-Salpeter kernel. The resulting tetraquark equations are expressed in an exact field theoretic form, and are in agreement with those obtained previously by consideration of disconnected interactions; however, despite being more general, they have been derived here in a much simpler and more transparent way.
\end{abstract}

\maketitle
\newpage

\section{Introduction}

In Quantum Field Theory (QFT) the number of particles is not conserved. This fact necessitates a careful theoretical definition of an exotic particle. In particular, to define a tetraquark,  an exotic bound state of two quarks and two antiquarks ($2q2\bar q$) whose existence has recently been evidenced \cite{Ablikim:2013mio,Liu:2013dau,Aaij:2020fnh}, requires more subtlety than to simply associate it with  a pole in the 4-body $2q2\bar q$ Green function $G^{(4)}$.  Indeed, in QFT the existence of a tetraquark is signalled  by a pole in  $G_{ir}^{(4)}$, the {\it $q\bar q$-irreducible part} of $G^{(4)}$, even though the physical mass of this tetraquark is determined by the corresponding pole position in the {\it full} Green function  $G^{(4)}$ (which will generally be shifted, perhaps slightly, with respect to the pole position in its $q\bar q$-irreducible part).
This fact is made clear in \fig{fig:G4} which demonstrates that {\em any} pole in the two-body $q\q$ Green function $\Gt$ will automatically appear in $\Gf$, making a pole in $\Gf$ an insufficient criterion for a tetraquark.

Similarly, if a candidate tetraquark pole is found in the $q\bar q$ Green function  $G^{(2)}$, a sign that it is indeed a tetraquark is the existence of a corresponding pole in $K^{(2)}_{4q-red}$, the  $2q2\bar q$ ($4q$)-reducible part of the two-body Bethe-Salpeter (BS) kernel $K^{(2)}$.
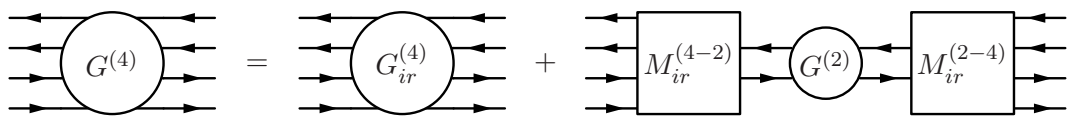
\begin{figure}[b]
\begin{center}
\begin{fmffile}{G4}
\fmfset{arrow_len}{2.5mm}
\begin{align*}
\parbox{27mm}{
\begin{fmfgraph*}(27,12)
\fmfstraight
\fmfleft{f4,f3,f2,f1}\fmfright{i4,i3,i2,i1}
\fmf{phantom,tension=1}{i1,r1,m1,l1,f1}
\fmf{phantom,tension=1}{i2,r2,m2,l2,f2}
\fmf{phantom,tension=1}{i3,r3,m3,l3,f3}
\fmf{phantom,tension=1}{i4,r4,m4,l4,f4}
\fmffreeze
\fmf{phantom,tension=1}{m4,c,m1}
\fmfv{d.sh=circle,d.f=empty,d.si=.5w, l.d=-0mm,l=$G^{(4)}$}{c}
\fmf{fermion}{l1,f1}\fmf{plain}{l1,r1}\fmf{fermion}{i1,r1}
\fmf{fermion}{l2,f2}\fmf{plain}{l2,r2}\fmf{fermion}{i2,r2}
\fmf{fermion}{f3,l3}\fmf{plain}{l3,r3}\fmf{fermion}{r3,i3}
\fmf{fermion}{f4,l4}\fmf{plain}{l4,r4}\fmf{fermion}{r4,i4}
\end{fmfgraph*}}
\hspace{3mm} &=\hspace{3mm}
\parbox{27mm}{
\begin{fmfgraph*}(27,12)
\fmfstraight
\fmfleft{f4,f3,f2,f1}\fmfright{i4,i3,i2,i1}
\fmf{phantom,tension=1}{i1,r1,m1,l1,f1}
\fmf{phantom,tension=1}{i2,r2,m2,l2,f2}
\fmf{phantom,tension=1}{i3,r3,m3,l3,f3}
\fmf{phantom,tension=1}{i4,r4,m4,l4,f4}
\fmffreeze
\fmf{phantom,tension=1}{m4,c,m1}
\fmfv{d.sh=circle,d.f=empty,d.si=.5w, l.d=-0mm,l=$G^{(4)}_{ir}$}{c}
\fmf{fermion}{l1,f1}\fmf{plain}{l1,r1}\fmf{fermion}{i1,r1}
\fmf{fermion}{l2,f2}\fmf{plain}{l2,r2}\fmf{fermion}{i2,r2}
\fmf{fermion}{f3,l3}\fmf{plain}{l3,r3}\fmf{fermion}{r3,i3}
\fmf{fermion}{f4,l4}\fmf{plain}{l4,r4}\fmf{fermion}{r4,i4}
\end{fmfgraph*}}
\hspace{3mm}+\hspace{3mm}
\parbox{27mm}{
\begin{fmfgraph*}(27,12)
\fmfstraight
\fmfleft{f4,f3,f2,f1}\fmfright{i4,i3,i2,i1}
\fmf{phantom,tension=1}{i1,r1,m1,l1,f1}
\fmf{phantom,tension=1}{i2,r2,m2,l2,f2}
\fmf{phantom,tension=1}{i3,r3,m3,l3,f3}
\fmf{phantom,tension=1}{i4,r4,m4,l4,f4}
\fmffreeze
\fmf{phantom,tension=1}{m4,c,m1}
\fmfv{d.sh=square,d.f=empty,d.si=.5w, l.d=-0mm,l=$M^{(4-2)}_{ir}$}{c}
\fmf{fermion}{l1,f1}\fmf{plain}{m1,l1}
\fmf{fermion}{l2,f2}\fmf{plain}{l2,r2}\fmf{fermion}{i2,r2}
\fmf{fermion}{f3,l3}\fmf{plain}{l3,r3}\fmf{fermion}{r3,i3}
\fmf{fermion}{f4,l4}\fmf{plain}{l4,m4}
\end{fmfgraph*}}
\hspace{-9.mm}
\parbox{27mm}{
\begin{fmfgraph*}(27,12)
\fmfstraight
\fmfleft{f4,f3,f2,f1}\fmfright{i4,i3,i2,i1}
\fmf{phantom,tension=1}{i1,r1,m1,l1,f1}
\fmf{phantom,tension=1}{i2,r2,m2,l2,f2}
\fmf{phantom,tension=1}{i3,r3,m3,l3,f3}
\fmf{phantom,tension=1}{i4,r4,m4,l4,f4}
\fmffreeze
\fmf{phantom,tension=1}{m4,c,m1}
\fmfv{d.sh=circle,d.f=empty,d.si=.35w, l.d=-0mm,l=$G^{(2)}$}{c}
\end{fmfgraph*}}
\hspace{-9.mm}
\parbox{27mm}{
\begin{fmfgraph*}(27,12)
\fmfstraight
\fmfleft{f4,f3,f2,f1}\fmfright{i4,i3,i2,i1}
\fmf{phantom,tension=1}{i1,r1,m1,l1,f1}
\fmf{phantom,tension=1}{i2,r2,m2,l2,f2}
\fmf{phantom,tension=1}{i3,r3,m3,l3,f3}
\fmf{phantom,tension=1}{i4,r4,m4,l4,f4}
\fmffreeze
\fmf{phantom,tension=1}{m4,c,m1}
\fmfv{d.sh=square,d.f=empty,d.si=.5w, l.d=-0mm,l=$M^{(2-4)}_{ir}$}{c}
\fmf{plain}{m1,r1}\fmf{fermion}{i1,r1}
\fmf{fermion}{l2,f2}\fmf{plain}{l2,r2}\fmf{fermion}{i2,r2}
\fmf{fermion}{f3,l3}\fmf{plain}{l3,r3}\fmf{fermion}{r3,i3}
\fmf{plain}{m4,r4}\fmf{fermion}{r4,i4}
\end{fmfgraph*}}
\end{align*}
\end{fmffile}   
\vspace{3mm}

\caption{\fign{fig:G4}  Field theoretic structure of the $2q2\q$ Green function $G^{(4)}$, where $G^{(4)}_{ir}$ is the $q\q$-irreducible part of $\Gf$, with $M^{(4-2)}_{ir}$  and $M^{(2-4)}_{ir}$ being $q\q$-irreducible $2q2 \leftarrow q\q$ and $q\q \leftarrow 2q2\q$ transition amplitudes, respectively.}
\end{center}
\end{figure}
These observations are particularly relevant to the recent efforts to describe tetraquarks using covariant few-body equations \cite{Heupel:2012ua,Kvinikhidze:2014yqa, Santowsky:2020pwd}.  The initial such  formulation  \cite{Heupel:2012ua} was based on an analysis of only the $q\bar q$-irreducible part of the $2q2\q$ Green function, $G^{(4)}_{ir}$ (although this fact was not emphasized at the time), and as coupling to $q\q$ channels was neglected, this analysis may not be the most accurate.
The present paper therefore addresses the more recent formulations of covariant few-body equations describing the four-body $2q 2\q$ system where coupling to two-body $q\q$ states is included \cite{Kvinikhidze:2014yqa, Santowsky:2020pwd}. The ultimate goal of such equations is to describe tetraquarks in terms of identical poles in the full $2q2\q$ and $q\q$ Green functions, $G^{(4)}$ and $G^{(2)}$, with both poles being due to a common pole in $G^{(4)}_{ir}$. Unfortunately there is currently no consensus on the form such equations take in the approximation where only two-body forces are retained  in the equation for $G^{(4)}_{ir}$
and where the underlying dynamics is dominated by meson-meson ($MM$) and diquark-antidiquark ($D\D$) components. 

A serious issue facing all relativistically covariant derivations of equations that couple four-body to two-body states, is the appearance of overcounted terms. This type of overcounting first came to light in the analogous problem of formulating covariant few-body equations for the pion-two-nucleon ($\pi N N $) system where coupling to two-nucleon  $(N N)$ states is included \cite{Kvinikhidze:1993bn}; there, it was found that in order to attain overcounting-free equations where all possible two-body interactions are retained, special subtraction terms needed to be introduced and certain three-body forces needed to be retained. Although covariant few-body equations for the coupled $q\q-2 q 2\q$ system can be derived in an analogous manner to that for the  $NN-\pi NN$ system, as  in Ref.\ \cite{Kvinikhidze:1993bn}, current interest is in formulating a less detailed approach that applies specifically to the case where  the $q\q-2q 2\q$ system is dominated by $MM$ and $D\D$ components. It is in this context that we would like to revisit the formulation of the coupled $q\q-2 q 2\q$ system.  In the process, we aim to help resolve the differences between the aforementioned tetraquark equations by presenting a derivation that does not involve the introduction of any  explicit disconnected contributions to two-body ($q\q$) interactions, viewed as potentially problematic in Ref.\ \cite{Santowsky:2020pwd}, but which were pivotal to our previous derivation \cite{Kvinikhidze:2014yqa}. Despite the very different approach presented here, the tetraquark equations resulting from the new derivation are in agreement with those of Ref.\ \cite{Kvinikhidze:2014yqa}, and moreover, are obtained in a simpler and much more transparent way. 

\section{Tetraquark poles and wave functions}

In the context of QFT, to formulate a few-body approach for a system of particles where some of them can be absorbed by others (e.g. $\pi$ by $N$ in the $\pi NN$ system or a $q\bar q$ pair that is annihilated in the $2q2\bar q$ system) one starts with the general structure of the full few-body Green function, which in the case of the $2q2\bar q$ system is manifested by the decomposition
\be
G^{(4)}=G^{(4)}_{ir}+G_{ir}^{(4-2)}G_0^{(2)-1}G^{(2)}G_0^{(2)-1}G_{ir}^{(2-4)}, \eqn{exact}
\ee
where $G_0^{(2)}$ is the disconnected part of the two-body $q\bar q$ Green function $G^{(2)}$  corresponding to the independent propagation of $q$ and $\q$ in the $s$ channel, and $G_{ir}^{(2-4)}$ ($G_{ir}^{(4-2)}$) is the sum of all
$q\bar q$-irreducible diagrams corresponding to the transition $q\bar q\leftarrow 2q2\bar q$ ($2q2\bar q\leftarrow q\bar q$). Equation (\ref{exact}) is illustrated in \fig{fig:G4} where $G_{ir}^{(4-2)}G_0^{(2)-1}\equiv M_{ir}^{(4-2)}$ and $G_0^{(2)-1}G_{ir}^{(2-4)}\equiv M_{ir}^{(2-4)}$.
The main problem is then to express $G_{ir}^{(2-4)}$ and $G_{ir}^{(4-2)}$ in terms of $G^{(4)}_{ir}$, while $G^{(4)}_{ir}$ can be expressed in terms of four-body scattering equations that are valid in the absence of $q\q$ annihilation.

As discussed above, the signature for the formation of a tetraquark in the $2q2\q$ system is the occurrence of a pole in $G^{(4)}_{ir}$. In the current problem setting, a pole in $G^{(4)}_{ir}$ is similarly the signature for the formation of a tetraquark bound state in the $q\q$ system. The contribution of  $G^{(4)}_{ir}$ to the $q\bar q$ Green function $G^{(2)}$ takes place through the BS kernel $K^{(2)}$ which, by definition, is $q\q$-irreducible and is related to $G^{(2)}$ through the Dyson equation\footnote{Note that our definition of $K^{(2)}$ is different from the one used in Ref.\ \cite{Santowsky:2020pwd}.}
\be
  G^{(2)} = G_0^{(2)}+G_0^{(2)}K^{(2)}G^{(2)}.   \eqn{G2}
\ee
In particular, the signature tetraquark bound state pole occurs in the $2q2\q$-reducible part of $K^{(2)}$, which can be expressed as 
\be
K^{(2)}_{4q-red} =  A^{(2-4)}G_{ir}^{(4)} A^{(4-2)}  \eqn{4q-red}
\ee
where $A^{(4-2)} $ ($A^{(2-4)} $) is some amplitude (of course $q\q$-irreducible)  corresponding to the transition  $2q2\bar q\leftarrow q\bar q$ ($q\bar q\leftarrow 2q2\bar q$).  The full analysis of these functions would be similar to the one used for the $\pi NN$ system in the covariant approach of Ref.\ \cite{Kvinikhidze:1993bn}, but this will be investigated elsewhere. In the present paper we use  the 4q-reducible part of the two-body potential,  \eq{4q-red}, to look for a tetraquark solution in the 2-body equation (of course if the potential itself possesses the corresponding pole at the tetraquark "bare" mass value); however, we follow the existing approaches where $MM$ and $D\bar D$ dominance is assumed.

Ultimately, we shall be interested in the case where $G^{(4)}$ and $G^{(2)}$ display simultaneous poles corresponding to a tetraquark of mass $M$, so that as $P^2\rightarrow M^2$ where $P$ is the total momentum of each system, 
\be
G^{(4)}\rightarrow i \frac{\Psi \bar\Psi}{P^2-M^2}, \hspace{1cm}
G^{(2)}\rightarrow i \frac{G_0^{(2)}\Gamma^* \bar\Gamma^*G_0^{(2)}}{P^2-M^2}.
\eqn{566}
\ee
In \eq{566}, $\Psi$ is the tetraquark  4-body $(2q2\q)$ bound state wave function, while $\Gamma^*$ is the form factor for the disintegration of a  tetraquark  into a $q\q$ pair. We note that the definition of $\Psi$ and $\Gamma^*$ via the pole parts of $G^{(4)}$ and $G^{(2)}$ in \eq{566}, together with \eq{exact} relating  $G^{(4)}$ and $G^{(2)}$, leads to the relation between $\Psi$ and
$\Gamma^*$,
\be
\Psi= G^{(4-2)} \Gamma^*. \eqn{exact-wf}
\ee

As is evident from \eq{G2} and the second of the relations in \eq{566}, a tetraquark state will also satisfy the two-body (not 4--body) equation,
\be
\Gamma^*=K^{(2)}G_0^{(2)}\Gamma^* .   \eqn{2b-tetr}
\ee
It is \eq{2b-tetr} which will be used in this paper to formulate the tetraquark equations. This will be achieved by first constructing $\Gf_{ir}$ using the four-body equations of Khvedelidze and Kvinikhidze \cite{Khvedelidze:1991qb}, together with a pole approximation Ansatz for all quark pair scattering amplitudes, and then using \eq{4q-red} to generate the essential part of the $q\q$ kernel. 

\section{Tetraquark few-body equations}

The approach used here to derive covariant equations for the coupled $q\q-2q2\q$ system is different from that employed by us in Ref.\  \cite{Kvinikhidze:2014yqa}. Instead of incorporating coupling to $q\q$ states right from the outset, as embodied in the full 4-body Green function  $\Gf$, here we first consider a formulation of 4-body tetraquark equations for the case where there is no coupling to $q\q$ states; that is, we first consider a formulation based on $\Gf_{ir}$, the $q\q$-irreducible part of $\Gf$. Coupling to $q\q$ states is then achieved by generating the $q\q$ kernel $K^{(2)}$ through a simple contraction of 4-body to 2-body states as in \eq{4q-red}. 

One can express $\Gf_{ir}$ in terms of the  $q\q$-irreducible 4-body interaction kernel $\Kf_{ir}$ through the Dyson equation
\begin{align}
\Gf_{ir}&=\Gf_0+\Gf_0\Kf_{ir}\Gf_{ir}
\end{align}
where $\Gf_0$ is the fully disconnected part of $\Gf$. For simplicity, we start out by treating the quarks as distinguishable particles; however, the full antisymmetry of quark states will be taken into account shortly. The kernel $\Kf_{ir}$ can be formally expressed as
\be
\Kf_{ir}=\Kf_{2F}+\Kf_{3F}
\ee
where $\Kf_{2F}$ consists of only pair-wise interactions, and $\Kf_{3F}$ consists of all other contributions, necessarily involving three- and four-body forces.  Assigning labels 1,2 to the quarks and 3,4 to the antiquarks, one can write $\Kf_{2F}$ as a sum of three terms whose structure is illustrated in \fig{K}, and correspondingly expressed as
\be
\Kf_{2F}=\sum_{aa'} \Kf_{aa'} =\sum_{\A} \Kf_{\A}  \eqn{pair}
\ee
where the index $a \in \left\{12,13,14,23,24,34\right\}$ enumerates six possible pairs of particles, 
the double index $aa'  \in \left\{(13,24), (14,23),(12,34)\right\}$ enumerates three possible two pairs of particles, and the Greek 
index $\A$ is used as an abbreviation for $aa'$ such that $\A=1$ denotes  $aa'=(13,24)$,  $\A=2$ denotes $aa'=(14,23)$, and  $\A=3$ denotes $aa'=(12,34)$.   Thus $\Kf_{aa'}$ describes the part of
the four-body kernel where all interactions are switched off except those within the pairs $a$ and $a'$.
\begin{figure}[t]
\begin{center}
\begin{fmffile}{K}
\[
\Kf_1=\hspace{7mm}
\parbox{20mm}{
\begin{fmfgraph*}(20,15)
\fmfstraight
\fmfleft{f4,f2,f3,f1}\fmfright{i4,i2,i3,i1}
\fmf{plain,tension=1.3}{i1,v1,f1}
\fmf{plain,tension=1.3}{i2,v2,f2}
\fmf{plain,tension=1.3}{i3,v3,f3}
\fmf{plain,tension=1.3}{i4,v4,f4}
\fmffreeze
\fmf{phantom,tension=1.3}{v1,v13,v3}
\fmfv{d.s=circle,d.f=empty,d.si=14,background=(1,,.51,,.5)}{v13}
\fmf{phantom,tension=1.3}{v2,v24,v4}
\fmfv{d.s=circle,d.f=empty,d.si=14,background=(1,,.51,,.5)}{v24}
\fmfv{label=$1$,l.a=0}{i1}
\fmfv{label=$2$,l.a=0}{i2}
\fmfv{label=$3$,l.a=0}{i3}
\fmfv{label=$4$,l.a=0}{i4}
\fmfv{label=$q$,l.a=180}{f1}
\fmfv{label=$q$,l.a=180}{f2}
\fmfv{label=$\q$,l.a=180}{f3}
\fmfv{label=$\q$,l.a=180}{f4}
\end{fmfgraph*}}\hspace{6mm},
\hspace{6mm}
\Kf_2 =
\hspace{7mm}
\parbox{20mm}{
\begin{fmfgraph*}(20,15)
\fmfstraight
\fmfleft{f3,f2,f4,f1}\fmfright{i3,i2,i4,i1}
\fmf{plain,tension=1.3}{i1,v1,f1}
\fmf{plain,tension=1.3}{i2,v2,f2}
\fmf{plain,tension=1.3}{i3,v3,f3}
\fmf{plain,tension=1.3}{i4,v4,f4}
\fmffreeze
\fmf{phantom,tension=1.3}{v1,v14,v4}
\fmfv{d.s=circle,d.f=empty,d.si=14,background=(1,,.51,,.5)}{v14}
\fmf{phantom,tension=1.3}{v2,v23,v3}
\fmfv{d.s=circle,d.f=empty,d.si=14,background=(1,,.51,,.5)}{v23}
\fmfv{label=$1$,l.a=0}{i1}
\fmfv{label=$2$,l.a=0}{i2}
\fmfv{label=$3$,l.a=0}{i3}
\fmfv{label=$4$,l.a=0}{i4}
\fmfv{label=$q$,l.a=180}{f1}
\fmfv{label=$q$,l.a=180}{f2}
\fmfv{label=$\q$,l.a=180}{f3}
\fmfv{label=$\q$,l.a=180}{f4}
\end{fmfgraph*}}
\hspace{6mm},\hspace{6mm}
\Kf_3 =
\hspace{7mm}
\parbox{20mm}{
\begin{fmfgraph*}(20,15)
\fmfstraight
\fmfleft{f4,f3,f2,f1}\fmfright{i4,i3,i2,i1}
\fmf{plain,tension=1.3}{i1,v1,f1}
\fmf{plain,tension=1.3}{i2,v2,f2}
\fmf{plain,tension=1.3}{i3,v3,f3}
\fmf{plain,tension=1.3}{i4,v4,f4}
\fmffreeze
\fmf{phantom,tension=1.3}{v1,v12,v2}
\fmfv{d.s=circle,d.f=empty,d.si=14,background=(.6235,,.7412,,1)}{v12}
\fmf{phantom,tension=1.3}{v3,v34,v4}
\fmfv{d.s=circle,d.f=empty,d.si=14,background=(.6235,,.7412,,1)}{v34}
\fmfv{label=$1$,l.a=0}{i1}
\fmfv{label=$2$,l.a=0}{i2}
\fmfv{label=$3$,l.a=0}{i3}
\fmfv{label=$4$,l.a=0}{i4}
\fmfv{label=$q$,l.a=180}{f1}
\fmfv{label=$q$,l.a=180}{f2}
\fmfv{label=$\q$,l.a=180}{f3}
\fmfv{label=$\q$,l.a=180}{f4}
\end{fmfgraph*}}
\]
\end{fmffile}   
\vspace{3mm}

\caption{\fign{K}  Structure of the terms $\Kf_\A$ ($\A=1, 2, 3$) making up the 4-body kernel $\Kf_{2F}$ where only two-body forces are included. The three terms are summed as in \eq{pair}. }
\end{center}
\end{figure}
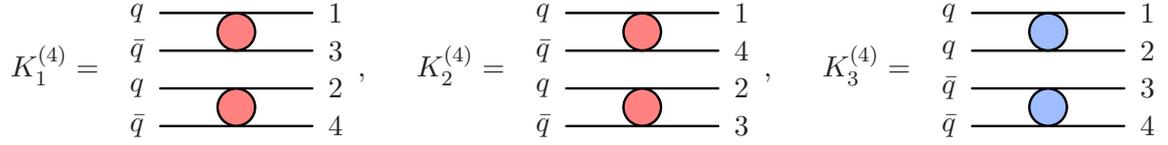
As is well known \cite{Khvedelidze:1991qb,Heupel:2012ua,Kvinikhidze:2014yqa},
$\Kf_{aa'}$ can be expressed in terms of the two-body kernels $K^{(2)}_a$ and $K^{(2)}_{a'}$ as
\be
K^{(4)}_{aa'}=K^{(2)}_a G_{a'}^{0}{}^{-1}+K^{(2)}_{a'}G_{a}^{0}{}^{-1} -K^{(2)}_aK^{(2)}_{a'}, \eqn{Kaa}
\ee
where $G_{a}^{0}$ ( $G_{a'}^{0}$ ) is the 2-body disconnected Green function for particle pair $a$ ($a'$).
It is also useful to introduce the corresponding 4-body $q\q$-irreducible t matrix $T_{ir}^{(4)}$ defined by equation
\begin{align}
\Gf_{ir}&=\Gf_0+\Gf_0T_{ir}^{(4)} \Gf_0.
\end{align}
One can similarly express  $T^{(4)}_{ir}$ as a sum of three terms \cite{Khvedelidze:1991qb}
\be
T^{(4)}_{ir}=\sum_{aa'} {\cal T}^{(4)}_{aa'} =\sum_{\A} {\cal T}^{(4)}_{\A}  \eqn{Tpair}
\ee
with components ${\cal T}^{(4)}_\A$ satisfying Faddeev-like equations
\be
{\cal T}^{(4)}_\A=T^{(4)}_\A  + \sum_\B T^{(4)}_\A  \bar\delta_{\A\B} G_0^{(4)} {\cal T}^{(4)}_\B \eqn{TD}
\ee
where $\bar\delta_{\A\B} = 1 - \delta_{\A\B}$ and where the Greek subscripts run over the three possible "two pairs" of particles as in \eq{pair}.
In \eq{TD}, $T^{(4)}_\A$ is the t matrix corresponding to kernel $\Kf_\A$, that is
\be
 T^{(4)}_\A = K^{(4)}_\A + K^{(4)}_\A \Gf_0 T^{(4)}_\A,
 \ee
with $T^{(4)}_\A$ being expressed in terms of  two-body t matrices $T^{(2)}_a$ and $T^{(2)}_{a'}$ as
\be
T^{(4)}_\A=T^{(4)}_{aa'}=T^{(2)}_a G_{a'}^{0}{}^{-1}+T^{(2)}_{a'}G_{a}^{0}{}^{-1} + T^{(2)}_a T^{(2)}_{a'}. \eqn{Taa}
\ee

\subsection{Tetraquark equations with no coupling to $q\q$ states}
To compare with the existing approaches \cite{Heupel:2012ua, Kvinikhidze:2014yqa,Santowsky:2020pwd}, our aim is to describe the tetraquark using two-body equations  that couple identical meson-meson ($MM$), and diquark-antidiquark ($D\D$) channels. To this end we consider $G_{ir}^{(4)}$ in the approximation
\begin{subequations} \eqn{assumptions}
\be
T^{(4)}_{aa'}=T^{(2)}_aT^{(2)}_{a'} \eqn{linear}
\ee
where the two-body t matrices $T^{(2)}_a$ and $T^{(2)}_{a'}$ are expressed in the bound state pole approximation
\be
T^{(2)}_a= i \Gamma_a D_a\bar\Gamma_a,
\ee 
\end{subequations}
where $D_a(P_a) = 1/(P_a^2-m_a^2)$ is the propagator for the bound particle (diquark, antidiquark, or meson) in the two-body channel $a$. Showing explicit dependence on momentum variables, $T^{(2)}_a$, for $a=12$, can be expressed as
$
T^{(2)}_{12}(p_1'p'_2,p_1p_2)=i\Gamma(p_1'p'_2)D_{12}(P)\bar\Gamma(p_1p_2),
$
where $P=p_1+p_2$ is the total off-mass-shell momentum of the bound particle.

As discussed above, the signature for a tetraquark is the existence of a pole in $\Gf_{ir}$. In turn, this means the existence of 
a 4-body tetraquark wave function $\Psi_{ir}\equiv \Gf_0 \psi$ for the case where all coupling to $q\q$ states is switched off. We therefore begin by considering the corresponding bound state form factor $\psi$ for the case of  2 indistinguishable quarks and 2 indistinguishable  antiquarks, and relate it to the corresponding form factor $\psi^d$ for distinguishable quarks as
\be
 \psi=\frac{1}{4}(1-{\cal P}_{12})  (1-{\cal P}_{34}) \psi^d  \eqn{id--und}
\ee
where ${\cal P}_{ij}$ is the operator exchanging the quantum numbers of particles $i$ and $j$.
The Faddeev-like equations for $\psi^d$ are \cite{Khvedelidze:1991qb,Heupel:2012ua},
\begin{subequations}
\begin{align}
\psi^d&=\sum_\A\psi^d_\A \\
\psi^d_\A &= \sum_\B T^{(4)}_\A  \bar\delta_{\A\B} G_0^{(4)}  \psi^d_{\B} \eqn{psiD}
\end{align}
\end{subequations}
where $\bar\delta_{\A\B} = 1 - \delta_{\A\B}$ and the Greek subscripts run over the three possible "two pairs" of particles as in \eq{pair}.
Using the  approximations of \eqs{assumptions}, one can write
\be
T^{(4)}_\A = T^{(2)}_a T^{(2)}_{a'} = i^2\Gamma_a\Gamma_{a'} D_a D_{a'}\bar\Gamma_a\bar\Gamma_{a'}\equiv -\Gamma_\A D_\A \bar\Gamma_\A
\ee
where $ \Gamma_\A\equiv \Gamma_a\Gamma_{a'} $,  $ \bar \Gamma_\A\equiv \bar \Gamma_a\bar \Gamma_{a'} $, and $D_\A\equiv D_a D_{a'} $. Further, defining the vertex functions $\phi^d_\A$ by the relation
\be
\psi^d_\A = \Gamma_\A D_\A \phi^d_\A,
\ee
it follows from \eq{psiD} that
\be
\phi^d_\A= \sum_\B V_{\A\B}D_\B \phi^d_\B    \eqn{Dtetra}
\ee
where
\be
V_{\A\B}=-\bar\delta_{\A\B}\bar\Gamma_\A G_0^{(4)}\Gamma_\B. 
\ee
Noting that $\Gamma_{12}=-\Gamma_{21}$ and $\Gamma_{34}=-\Gamma_{43}$, it follows  that $V_{12}=V_{21}$,  $V_{23}=-V_{13}$ and $V_{32}=-V_{31}$.

We can now use \eq{id--und} to derive $MM$ and $D\bar D$ components of the tetraquark form factor $\psi$  in the case of indistinguishable quarks. These are defined by  the pole contributions to $\psi$ at 
$p^2_{13}=M_\pi^2$, $p^2_{24}=M_\pi^2$, $p^2_{14}=M_\pi^2$, $p^2_{23}=M_\pi^2$,
$p^2_{12}=M_D^2 $, and $p^2_{34}=M_D^2$, where $p_{ij}=p_i+p_j$ is the total momentum of particles $i$ and $j$, $M_\pi$ is the mass of the meson and $M_D$ is the mass of the diquark or antidiquark. To this end consider  the use of \eq{Dtetra} in \eq{id--und}:
\begin{align}
 \psi&=\frac{1}{4}(1-{\cal P}_{14})  (1-{\cal P}_{34}) \left[\Gamma_1D_1\phi^d_1(p_{13},p_{24})+\Gamma_2D_2\phi^d_2(p_{14},p_{23})+\Gamma_3D_3\phi^d_3(p_{12},p_{34})\right]   \nn[3mm]
&
=\frac{1}{4}
\Gamma_{13}\Gamma_{24}D_1\left[\phi^d_1(p_{13},p_{24})+\phi^d_1(p_{24},p_{13})\right]-\frac{1}{4}\Gamma_{14}\Gamma_{23}D_2\left[\phi^d_1(p_{23},p_{14}) +\phi^d_1(p_{14},p_{23})\right]  \nn
&
+\frac{1}{4}\Gamma_{14}\Gamma_{23}D_2\left[\phi^d_2(p_{14},p_{23}) +\phi^d_2(p_{23},p_{14})\right] 
-\frac{1}{4}\Gamma_{13}\Gamma_{24}D_1\left[\phi^d_2(p_{24},p_{13}) +\phi^d_2( p_{13},p_{24})\right]  \nn
& +\Gamma_{12}\Gamma_{34}D_3\phi^d_3(p_{12},p_{34})      \nn[3mm]
& =\frac{1}{2}
\Gamma_{13}\Gamma_{24}D_1\left[\phi^S_1(p_{13},p_{24})- \phi^S_2(p_{13},p_{24})\right]  
+\frac{1}{2}\Gamma_{14}\Gamma_{23}D_2\left[\phi^d_2(p_{14},p_{23}) -\phi^d_1( p_{14},p_{23})\right]  \nn
&
+\Gamma_{12}\Gamma_{34}D_3\phi^d_3(p_{12},p_{34})  \nn[3mm]
&
=\frac{1}{2}\Gamma_{13}\Gamma_{24}D_1\phi_M(p_{13},p_{24})
-\frac{1}{2}\Gamma_{14}\Gamma_{23}D_2\phi_M(p_{14},p_{23}) +\Gamma_{12}\Gamma_{34}D_3\phi_D(p_{12},p_{34}),  \eqn{ind-dist}
\end{align}
where
\begin{subequations}\eqn{defPS}
\begin{align}
\phi^S_1(p,q)& =\frac{1}{2} \left[\phi^d_1(p,q) +\phi^d_1(q,p )\right],  \\
\phi^S_2(p,q) &=\frac{1}{2} \left[\phi^d_2(p,q) +\phi^d_2(q,p )\right].  
\end{align}
\end{subequations}
are symmetric functions under the exchange of the meson quantum numbers, and
\begin{subequations}\eqn{defMD}
\begin{align}
\phi_M(p,q)& = \phi^S_1(p,q) -\phi^S_2(p,q ),   \\
\phi_D(p,q) &= \phi^d_3(p,q)  
\end{align}
\end{subequations}
define the $MM$ and $D\bar D$ components of the tetraquark form factor $\psi$ where quarks are identical. 

To derive equations for the tetraquark vertex functions for identical quarks, we first write out \eq{Dtetra} for distinguishable quarks 
using notation  $V_{13}=V_{1D}$, $V_{23}=V_{2D}$, $V_{31}=V_{D1}$, $V_{32}=V_{D2}$, and $\phi_D=\phi^d_{12,34}$:
\begin{subequations} \eqn{1=1324*}
\begin{align}
\phi^d_1&=V_{12}D_2\phi^d_2+V_{1D}D_3\phi_D   \\
\phi^d_2&=V_{21}D_1\phi^d_1+V_{2D}D_3\phi_D =V_{12}D_1\phi^d_1-V_{1D}D_3\phi_D \\
\phi_D&=V_{D1}D_1\phi^d_1+V_{D2}D_2\phi^d_2=V_{D1}(D_1\phi^d_1-D_2\phi^d_2 ). 
\end{align}
\end{subequations}
Then, subtracting the second line from the first, we obtain a set of two equations for $\phi_-=\phi^d_1-\phi^d_2$ and $\phi_D$, 
\begin{subequations}\eqn{PhiMD}
\begin{align}
\phi_-&=-V_{12}D_2\phi_-+2V_{1D}D_3\phi_D \eqn{PhiMDa} \nn
&=-2V_{12}\left(\frac{1}{2}MM\right)\phi_-+2V_{1D}D\bar D\phi_D ,\\
\phi_D&= V_{D1}D_1\phi_- = 2V_{D1}\left(\frac{1}{2}MM\right)\phi_- 
\end{align}
\end{subequations}
where we used $D_1=D_2\equiv  MM $, $D_3\equiv D\bar D$.
Equations (\ref{PhiMD}) can be written in matrix form as 
\be
\left( \begin{array}{c} \phi_-\\
 \phi_D \end{array} \right)
=2 \left( \begin{array}{cc}- V_{12}
& V_{1D}\\
 V_{D1} &0 \end{array} \right) 
 \left( \begin{array}{cc}\frac{1}{2}MM & 0 \\
 0 &D\bar D \end{array} \right) 
 \left( \begin{array}{c} \phi_-\\
 \phi_D \end{array} \right)    .
\ee
To finally  derive the tetraquark equations in the case of indistinguishable quarks,  note that according \eqs{defPS},
\begin{align}
\phi_M&= \Phi^S_1-\Phi^S_2=\Phi^S_-=\frac{1}{2} \left[\phi_-(p,q) +\phi_-(q,p )\right]\nn
&=\frac{1}{2}(1+{\cal P}) \phi_-
\end{align}
where ${\cal P}$ is permutation operator of the meson  state  labels. Thus, symmetrizing \eqs{PhiMD} with respect to meson legs gives
\begin{subequations}\eqn{PhiMD-S}
\begin{align}
\phi_M&  =-2\frac{1}{2}(1+{\cal P})V_{12}\left(\frac{1}{2}MM\right)\phi_-+2\frac{1}{2}(1+{\cal P})V_{1D}D\bar D\phi_D ,  \nn
&
=-2\frac{1}{2}(1+{\cal P})V_{12}\left(\frac{1}{2}MM\right)\phi_M+2V_{1D}D\bar D\phi_D , \\
\phi_D& = 2V_{D1}\left(\frac{1}{2}MM\right)\phi_-= 2V_{D1}\left(\frac{1}{2}MM\right)\phi_M , 
\end{align}
\end{subequations}
where we have used the following symmetry properties of $V_{12}$ and $V_{1D}$:
\begin{subequations}
\begin{align}
(1+{\cal P})V_{12} &=(1+{\cal P})V_{12}\frac{1}{2}(1+{\cal P}), \\
\frac{1}{2}(1+{\cal P})V_{1D}&=V_{1D}.
\end{align}
\end{subequations}
Equations (\ref{PhiMD-S}) can be written in matrix form as
\be
\phi = V G_0^M \phi    \eqn{id-q-tet}
\ee
where
\be
\phi = 
\begin{pmatrix}
\phi_M \\ \phi_D 
\end{pmatrix}
,\hspace{1cm}
G_0^M=
\begin{pmatrix}
\frac{1}{2} MM & 0 \\ 
0 & D\bar D 
\end{pmatrix},
\ee
and
\be
V = \begin{pmatrix}
-(1+{\cal P})V_{12} & 2V_{1D}\\
2 V_{D1} & 0 
\end{pmatrix},  \eqn{Vmat}
\ee
thereby revealing $V$  of \eq{Vmat} to be the interaction kernel for the coupled $MM-D\bar D$ system. The elements of $V$ involve the potentials
\begin{subequations} \eqn{Nota-Vij}
\begin{align}
V_{12}&=-\bar\Gamma_1G_0^{(4)}\Gamma_2 =-\bar\Gamma_{13} \bar\Gamma_{24}G_0^{(4)}\Gamma_{14} \Gamma_{23}, \\
V_{1D}&=-\bar\Gamma_1G_0^{(4)}\Gamma_3= -\bar\Gamma_{13} \bar\Gamma_{24}G_0^{(4)}\Gamma_{12} \Gamma_{34}, \\
V_{D1}&=-\bar\Gamma_3G_0^{(4)}\Gamma_1= -\bar\Gamma_{12} \bar\Gamma_{34}G_0^{(4)}\Gamma_{13} \Gamma_{24} ,  \end{align}
\end{subequations}
as illustrated in \fig{pots}. 
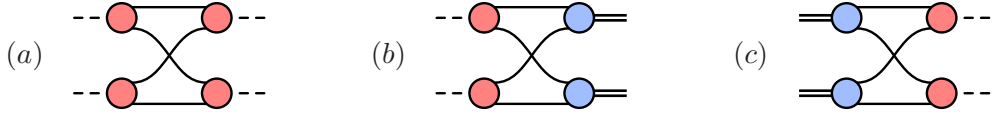
\begin{figure}[t]
\begin{center}
\begin{fmffile}{V}
\begin{align*}
(a)\hspace{4mm}
\parbox{25mm}{
\begin{fmfgraph*}(25,15)
\fmfstraight
\fmfleftn{l}{7}\fmfrightn{r}{7}\fmfbottomn{b}{5}\fmftopn{t}{5}
\fmf{phantom}{l2,vb1,mb1,vb2,r2}
\fmf{phantom}{l6,vt1,mt1,vt2,r6}
\fmf{phantom}{l2,xb1,mb1,xb2,r2}
\fmf{phantom}{l6,xt1,mt1,xt2,r6}
\fmf{phantom}{l2,yb1,mb1,yb2,r2}
\fmf{phantom}{l6,yt1,mt1,yt2,r6}
\fmffreeze
\fmfshift{4 right}{xb1}
\fmfshift{4 down}{xb1}
\fmfshift{4 left}{xt2}
\fmfshift{4 down}{xt2}
\fmfshift{4 right}{xt1}
\fmfshift{4 down}{xt1}
\fmfshift{4 left}{xb2}
\fmfshift{4 down}{xb2}
\fmfshift{4 right}{yb1}
\fmfshift{4 up}{yb1}
\fmfshift{4 left}{yt2}
\fmfshift{4 up}{yt2}
\fmfshift{4 right}{yt1}
\fmfshift{4 up}{yt1}
\fmfshift{4 left}{yb2}
\fmfshift{4 up}{yb2}
\fmfv{d.sh=circle,d.f=empty,d.si=11,background=(1,,.51,,.5)}{vb1}
\fmfv{d.sh=circle,d.f=empty,d.si=11,background=(1,,.51,,.5)}{vb2}
\fmfv{d.sh=circle,d.f=empty,d.si=11,background=(1,,.51,,.5)}{vt1}
\fmfv{d.sh=circle,d.f=empty,d.si=11,background=(1,,.51,,.5)}{vt2}
\fmf{dashes}{l2,vb1}
\fmf{dashes}{l6,vt1}
\fmf{dashes}{r2,vb2}
\fmf{dashes}{r6,vt2}
\fmfi{plain}{vloc(__xt2) {left}..tension 1 ..{left}vloc(__yb1)}
\fmfi{plain}{vloc(__xt1) {right}.. tension 1 ..{right}vloc(__yb2)}
\fmfi{plain}{vloc(__yt2) .. vloc(__yt1)}
\fmfi{plain}{vloc(__xb1) .. vloc(__xb2)}
\end{fmfgraph*}}
\hspace{14mm} (b) \hspace{4mm}
\parbox{25mm}{
\begin{fmfgraph*}(25,15)
\fmfstraight
\fmfleftn{l}{7}\fmfrightn{r}{7}\fmfbottomn{b}{5}\fmftopn{t}{5}
\fmf{phantom}{l2,vb1,mb1,vb2,r2}
\fmf{phantom}{l6,vt1,mt1,vt2,r6}
\fmf{phantom}{l2,xb1,mb1,xb2,r2}
\fmf{phantom}{l6,xt1,mt1,xt2,r6}
\fmf{phantom}{l2,yb1,mb1,yb2,r2}
\fmf{phantom}{l6,yt1,mt1,yt2,r6}
\fmffreeze
\fmfshift{4 right}{xb1}
\fmfshift{4 down}{xb1}
\fmfshift{4 left}{xt2}
\fmfshift{4 down}{xt2}
\fmfshift{4 right}{xt1}
\fmfshift{4 down}{xt1}
\fmfshift{4 left}{xb2}
\fmfshift{4 down}{xb2}
\fmfshift{4 right}{yb1}
\fmfshift{4 up}{yb1}
\fmfshift{4 left}{yt2}
\fmfshift{4 up}{yt2}
\fmfshift{4 right}{yt1}
\fmfshift{4 up}{yt1}
\fmfshift{4 left}{yb2}
\fmfshift{4 up}{yb2}
\fmfv{d.sh=circle,d.f=empty,d.si=11,background=(1,,.51,,.5)}{vb1}
\fmfv{d.sh=circle,d.f=empty,d.si=11,background=(.6235,,.7412,,1)}{vb2}
\fmfv{d.sh=circle,d.f=empty,d.si=11,background=(1,,.51,,.5)}{vt1}
\fmfv{d.sh=circle,d.f=empty,d.si=11,background=(.6235,,.7412,,1)}{vt2}
\fmf{dashes}{l2,vb1}
\fmf{dashes}{l6,vt1}
\fmf{dbl_plain}{r2,vb2}
\fmf{dbl_plain}{r6,vt2}
\fmfi{plain}{vloc(__xt2) {left}..tension 1 ..{left}vloc(__yb1)}
\fmfi{plain}{vloc(__xt1) {right}.. tension 1 ..{right}vloc(__yb2)}
\fmfi{plain}{vloc(__yt2) .. vloc(__yt1)}
\fmfi{plain}{vloc(__xb1) .. vloc(__xb2)}
\end{fmfgraph*}}
\hspace{14mm}(c) \hspace{4mm}
\parbox{25mm}{
\begin{fmfgraph*}(25,15)
\fmfstraight
\fmfleftn{l}{7}\fmfrightn{r}{7}\fmfbottomn{b}{5}\fmftopn{t}{5}
\fmf{phantom}{l2,vb1,mb1,vb2,r2}
\fmf{phantom}{l6,vt1,mt1,vt2,r6}
\fmf{phantom}{l2,xb1,mb1,xb2,r2}
\fmf{phantom}{l6,xt1,mt1,xt2,r6}
\fmf{phantom}{l2,yb1,mb1,yb2,r2}
\fmf{phantom}{l6,yt1,mt1,yt2,r6}
\fmffreeze
\fmfshift{4 right}{xb1}
\fmfshift{4 down}{xb1}
\fmfshift{4 left}{xt2}
\fmfshift{4 down}{xt2}
\fmfshift{4 right}{xt1}
\fmfshift{4 down}{xt1}
\fmfshift{4 left}{xb2}
\fmfshift{4 down}{xb2}
\fmfshift{4 right}{yb1}
\fmfshift{4 up}{yb1}
\fmfshift{4 left}{yt2}
\fmfshift{4 up}{yt2}
\fmfshift{4 right}{yt1}
\fmfshift{4 up}{yt1}
\fmfshift{4 left}{yb2}
\fmfshift{4 up}{yb2}
\fmfv{d.sh=circle,d.f=empty,d.si=11,background=(1,,.51,,.5)}{vb2}
\fmfv{d.sh=circle,d.f=empty,d.si=11,background=(.6235,,.7412,,1)}{vb1}
\fmfv{d.sh=circle,d.f=empty,d.si=11,background=(1,,.51,,.5)}{vt2}
\fmfv{d.sh=circle,d.f=empty,d.si=11,background=(.6235,,.7412,,1)}{vt1}
\fmf{dashes}{r2,vb2}
\fmf{dashes}{r6,vt2}
\fmf{dbl_plain}{l2,vb1}
\fmf{dbl_plain}{l6,vt1}
\fmfi{plain}{vloc(__xt2) {left}..tension 1 ..{left}vloc(__yb1)}
\fmfi{plain}{vloc(__xt1) {right}.. tension 1 ..{right}vloc(__yb2)}
\fmfi{plain}{vloc(__yt2) .. vloc(__yt1)}
\fmfi{plain}{vloc(__xb1) .. vloc(__xb2)}
\end{fmfgraph*}}
\end{align*}
\end{fmffile}   
\vspace{3mm}

\caption{\fign{pots}  The potentials making up the elements of the coupled channel $MM-D\bar D$ kernel matrix $V$ of \eq{Vmat}: (a) $V_{12}$, (b) $V_{1D}$, and (c) $V_{D1}$. Solid lines represent quarks or antiquarks, dashed lines represent mesons, and double-lines represent diquarks and antidiquarks.}
\end{center}
\end{figure}
With the kernel matrix $V$ established, one can determine the t matrix $T$ defined by 
\be
T= V + V G_0^M T ,  \eqn{T-MD}
\ee
and thereafter $G_0^M + G_0^M T G_0^M$, which is the matrix Green function in coupled $MM$-$D\bar D$ space corresponding to  $\Gf_{ir}$. As indicated by \eq{4q-red}, the $4q$-reducible part of the two-body $q\q$ kernel, $K^{(2)}_{4q-red}$, can be found by sandwiching $\Gf_{ir}$ between amplitudes that contract $2q2\q$  states to $q\q$ states. In the present case of coupled $MM$-$D\bar D$ channels, this contraction can be expressed as
\be
K^{(2)}_{4q-red} = \bar N (G_0^M + G_0^M T G_0^M) N
\ee
where $\bar N=(\bar N_M,\bar N_D)$ is the two-component amplitude whose elements  $\bar N_M$ and $\bar N_D$ describe transitions of two-meson and diquark-antidiquark states to the quark-antiquark state, $q\bar q\leftarrow M(p)M(k)$ and $q\bar q\leftarrow D(p)\bar D(k)$, respectively. Similarly,  $N=( N_M, N_D)$ describes transitions $M(p)M(k) \leftarrow q\bar q$ and $ D(p)\bar D(k) \leftarrow q\bar q$. Explicitly, these transition amplitudes are given by
\begin{subequations}  \eqn{Ns}
\begin{align} 
\bar N_M&=S_{23}(\Gamma_{13}^{p} \Gamma_{24}^{k}+\Gamma_{13}^{k} \Gamma_{24}^{p}),\hspace{5mm}
\bar N_D=S_{23}\Gamma_{12}^{p} \Gamma_{34}^{k} ,\\[2mm]
N_M&= (\bar\Gamma_{13}^p \bar\Gamma_{24}^k+\bar\Gamma_{13}^k \bar\Gamma_{24}^p)S_{23},\hspace{5mm}
N_D= \bar\Gamma_{12}^p \bar\Gamma_{34}^k S_{23},
\end{align}
\end{subequations}
where $S_{23}$ is the quark propagator connecting quark lines 2 and 3. Equations (\ref{Ns}) are illustrated in \fig{fig:Ns}.
\begin{figure}[t]
\begin{center}
\begin{fmffile}{N}
\begin{align*}
\bar N_M & \hspace{1mm} =\hspace{1mm} 
\parbox{15mm}{
\begin{fmfgraph*}(15,15)
\fmfstraight
\fmfleftn{l}{7}\fmfrightn{r}{7}\fmfbottomn{b}{4}\fmftopn{t}{4}
\fmf{phantom}{l2,mb1,vb2,r2}
\fmf{phantom}{l6,mt1,vt2,r6}
\fmf{phantom}{l2,mb1,xb2,r2}
\fmf{phantom}{l6,mt1,xt2,r6}
\fmf{phantom}{l2,mb1,yb2,r2}
\fmf{phantom}{l6,mt1,yt2,r6}
\fmf{phantom,tension=3}{l2,mb1}
\fmf{phantom,tension=3}{l6,mt1}
\fmffreeze
\fmfshift{4 down}{l2}
\fmfshift{4 up}{l6}
\fmfshift{4 left}{xt2}
\fmfshift{4 down}{xt2}
\fmfshift{4 left}{xb2}
\fmfshift{4 down}{xb2}
\fmfshift{4 left}{yt2}
\fmfshift{4 up}{yt2}
\fmfshift{4 left}{yb2}
\fmfshift{4 up}{yb2}
\fmfv{d.sh=circle,d.f=empty,d.si=11,background=(1,,.51,,.5)}{vb2}
\fmfv{d.sh=circle,d.f=empty,d.si=11,background=(1,,.51,,.5)}{vt2}
\fmf{dashes}{r2,vb2}
\fmfv{l=$p$,l.a=0,l.d=.07w}{r6}
\fmfv{l=$k$,l.a=0,l.d=.07w}{r2}
\fmf{dashes}{r6,vt2}
\fmfi{plain}{vloc(__xt2) {left}..tension 1 ..{right}vloc(__yb2)}
\fmfi{plain}{vloc(__yt2) .. vloc(__l6)}
\fmfi{plain}{vloc(__l2) .. vloc(__xb2)}
\end{fmfgraph*}}
\hspace{5mm}+\hspace{2mm}
\parbox{15mm}{
\begin{fmfgraph*}(15,15)
\fmfstraight
\fmfleftn{l}{7}\fmfrightn{r}{7}\fmfbottomn{b}{4}\fmftopn{t}{4}
\fmf{phantom}{l2,mb1,vb2,r2}
\fmf{phantom}{l6,mt1,vt2,r6}
\fmf{phantom}{l2,mb1,xb2,r2}
\fmf{phantom}{l6,mt1,xt2,r6}
\fmf{phantom}{l2,mb1,yb2,r2}
\fmf{phantom}{l6,mt1,yt2,r6}
\fmf{phantom,tension=3}{l2,mb1}
\fmf{phantom,tension=3}{l6,mt1}
\fmffreeze
\fmfshift{4 down}{l2}
\fmfshift{4 up}{l6}
\fmfshift{4 left}{xt2}
\fmfshift{4 down}{xt2}
\fmfshift{4 left}{xb2}
\fmfshift{4 down}{xb2}
\fmfshift{4 left}{yt2}
\fmfshift{4 up}{yt2}
\fmfshift{4 left}{yb2}
\fmfshift{4 up}{yb2}
\fmfv{d.sh=circle,d.f=empty,d.si=11,background=(1,,.51,,.5)}{vb2}
\fmfv{d.sh=circle,d.f=empty,d.si=11,background=(1,,.51,,.5)}{vt2}
\fmf{dashes}{r2,vb2}
\fmfv{l=$k$,l.a=0,l.d=.07w}{r6}
\fmfv{l=$p$,l.a=0,l.d=.07w}{r2}
\fmf{dashes}{r6,vt2}
\fmfi{plain}{vloc(__xt2) {left}..tension 1 ..{right}vloc(__yb2)}
\fmfi{plain}{vloc(__yt2) .. vloc(__l6)}
\fmfi{plain}{vloc(__l2) .. vloc(__xb2)}
\end{fmfgraph*}}\hspace{5mm},
\hspace{15mm}  
\bar N_D  \hspace{1mm} =\hspace{1mm} 
\parbox{15mm}{
\begin{fmfgraph*}(15,15)
\fmfstraight
\fmfleftn{l}{7}\fmfrightn{r}{7}\fmfbottomn{b}{4}\fmftopn{t}{4}
\fmf{phantom}{l2,mb1,vb2,r2}
\fmf{phantom}{l6,mt1,vt2,r6}
\fmf{phantom}{l2,mb1,xb2,r2}
\fmf{phantom}{l6,mt1,xt2,r6}
\fmf{phantom}{l2,mb1,yb2,r2}
\fmf{phantom}{l6,mt1,yt2,r6}
\fmf{phantom,tension=3}{l2,mb1}
\fmf{phantom,tension=3}{l6,mt1}
\fmffreeze
\fmfshift{4 down}{l2}
\fmfshift{4 up}{l6}
\fmfshift{4 left}{xt2}
\fmfshift{4 down}{xt2}
\fmfshift{4 left}{xb2}
\fmfshift{4 down}{xb2}
\fmfshift{4 left}{yt2}
\fmfshift{4 up}{yt2}
\fmfshift{4 left}{yb2}
\fmfshift{4 up}{yb2}
\fmfv{d.sh=circle,d.f=empty,d.si=11,background=(.6235,,.7412,,1)}{vb2}
\fmfv{d.sh=circle,d.f=empty,d.si=11,background=(.6235,,.7412,,1)}{vt2}
\fmf{dbl_plain}{r2,vb2}
\fmf{dbl_plain}{r6,vt2}
\fmfv{l=$p$,l.a=0,l.d=.07w}{r6}
\fmfv{l=$k$,l.a=0,l.d=.07w}{r2}
\fmfi{plain}{vloc(__xt2) {left}..tension 1 ..{right}vloc(__yb2)}
\fmfi{plain}{vloc(__yt2) .. vloc(__l6)}
\fmfi{plain}{vloc(__l2) .. vloc(__xb2)}
\end{fmfgraph*}}\\[5mm]
 N_M & \hspace{1mm} =\hspace{4mm} 
\parbox{15mm}{
\begin{fmfgraph*}(15,15)
\fmfstraight
\fmfleftn{l}{7}\fmfrightn{r}{7}\fmfbottomn{b}{4}\fmftopn{t}{4}
\fmf{phantom}{r2,mb1,vb2,l2}
\fmf{phantom}{r6,mt1,vt2,l6}
\fmf{phantom}{r2,mb1,xb2,l2}
\fmf{phantom}{r6,mt1,xt2,l6}
\fmf{phantom}{r2,mb1,yb2,l2}
\fmf{phantom}{r6,mt1,yt2,l6}
\fmf{phantom,tension=3}{r2,mb1}
\fmf{phantom,tension=3}{r6,mt1}
\fmffreeze
\fmfshift{4 down}{r2}
\fmfshift{4 up}{r6}
\fmfshift{4 right}{xt2}
\fmfshift{4 down}{xt2}
\fmfshift{4 right}{xb2}
\fmfshift{4 down}{xb2}
\fmfshift{4 right}{yt2}
\fmfshift{4 up}{yt2}
\fmfshift{4 right}{yb2}
\fmfshift{4 up}{yb2}
\fmfv{d.sh=circle,d.f=empty,d.si=11,background=(1,,.51,,.5)}{vb2}
\fmfv{d.sh=circle,d.f=empty,d.si=11,background=(1,,.51,,.5)}{vt2}
\fmf{dashes}{l2,vb2}
\fmfv{l=$p$,l.a=180,l.d=.07w}{l6}
\fmfv{l=$k$,l.a=180,l.d=.07w}{l2}
\fmf{dashes}{l6,vt2}
\fmfi{plain}{vloc(__xt2) {right}..tension 1 ..{left}vloc(__yb2)}
\fmfi{plain}{vloc(__yt2) .. vloc(__r6)}
\fmfi{plain}{vloc(__r2) .. vloc(__xb2)}
\end{fmfgraph*}}
\hspace{2mm}+\hspace{5mm}
\parbox{15mm}{
\begin{fmfgraph*}(15,15)
\fmfstraight
\fmfleftn{l}{7}\fmfrightn{r}{7}\fmfbottomn{b}{4}\fmftopn{t}{4}
\fmf{phantom}{r2,mb1,vb2,l2}
\fmf{phantom}{r6,mt1,vt2,l6}
\fmf{phantom}{r2,mb1,xb2,l2}
\fmf{phantom}{r6,mt1,xt2,l6}
\fmf{phantom}{r2,mb1,yb2,l2}
\fmf{phantom}{r6,mt1,yt2,l6}
\fmf{phantom,tension=3}{r2,mb1}
\fmf{phantom,tension=3}{r6,mt1}
\fmffreeze
\fmfshift{4 down}{r2}
\fmfshift{4 up}{r6}
\fmfshift{4 right}{xt2}
\fmfshift{4 down}{xt2}
\fmfshift{4 right}{xb2}
\fmfshift{4 down}{xb2}
\fmfshift{4 right}{yt2}
\fmfshift{4 up}{yt2}
\fmfshift{4 right}{yb2}
\fmfshift{4 up}{yb2}
\fmfv{d.sh=circle,d.f=empty,d.si=11,background=(1,,.51,,.5)}{vb2}
\fmfv{d.sh=circle,d.f=empty,d.si=11,background=(1,,.51,,.5)}{vt2}
\fmf{dashes}{l2,vb2}
\fmfv{l=$k$,l.a=180,l.d=.07w}{l6}
\fmfv{l=$p$,l.a=180,l.d=.07w}{l2}
\fmf{dashes}{l6,vt2}
\fmfi{plain}{vloc(__xt2) {right}..tension 1 ..{left}vloc(__yb2)}
\fmfi{plain}{vloc(__yt2) .. vloc(__r6)}
\fmfi{plain}{vloc(__r2) .. vloc(__xb2)}
\end{fmfgraph*}}\hspace{2mm},
\hspace{15mm}  
N_D  \hspace{1mm} =\hspace{4mm} 
\parbox{15mm}{
\begin{fmfgraph*}(15,15)
\fmfstraight
\fmfleftn{l}{7}\fmfrightn{r}{7}\fmfbottomn{b}{4}\fmftopn{t}{4}
\fmf{phantom}{r2,mb1,vb2,l2}
\fmf{phantom}{r6,mt1,vt2,l6}
\fmf{phantom}{r2,mb1,xb2,l2}
\fmf{phantom}{r6,mt1,xt2,l6}
\fmf{phantom}{r2,mb1,yb2,l2}
\fmf{phantom}{r6,mt1,yt2,l6}
\fmf{phantom,tension=3}{r2,mb1}
\fmf{phantom,tension=3}{r6,mt1}
\fmffreeze
\fmfshift{4 down}{r2}
\fmfshift{4 up}{r6}
\fmfshift{4 right}{xt2}
\fmfshift{4 down}{xt2}
\fmfshift{4 right}{xb2}
\fmfshift{4 down}{xb2}
\fmfshift{4 right}{yt2}
\fmfshift{4 up}{yt2}
\fmfshift{4 right}{yb2}
\fmfshift{4 up}{yb2}
\fmfv{d.sh=circle,d.f=empty,d.si=11,background=(.6235,,.7412,,1)}{vb2}
\fmfv{d.sh=circle,d.f=empty,d.si=11,background=(.6235,,.7412,,1)}{vt2}
\fmf{dbl_plain}{l2,vb2}
\fmf{dbl_plain}{l6,vt2}
\fmfv{l=$p$,l.a=180,l.d=.07w}{l6}
\fmfv{l=$k$,l.a=180,l.d=.07w}{l2}
\fmfi{plain}{vloc(__xt2) {right}..tension 1 ..{left}vloc(__yb2)}
\fmfi{plain}{vloc(__yt2) .. vloc(__r6)}
\fmfi{plain}{vloc(__r2) .. vloc(__xb2)}
\end{fmfgraph*}}
\end{align*}
\end{fmffile}   
\vspace{3mm}

\caption{\fign{fig:Ns}  Illustration of \eqs{Ns}. Lines have the same meaning as in \fig{pots}.}
\end{center}
\end{figure}
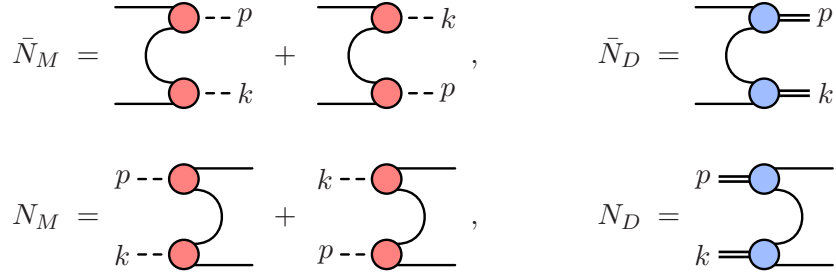

Using the formal solution to \eq{T-MD},
\be
T=\left(1-V G_0^M\right)^{-1}G_0^{M-1} -G_0^{M-1} ,   \eqn{*-*K*}
\ee
we can write the general expression for the two-body $q\q$ kernel as
\begin{align}
K^{(2)}&=\Delta + K^{(2)}_{4q-red}\nn
&= \Delta +\bar NG_0^M\left(1-VG_0^M\right)^{-1}N \eqn{488*}
\end{align}
where  $\Delta$  is defined to be the sum of all $q\bar q$-irreducible contributions allowed by QFT that are not accounted for by the last term of of \eq{488*}. In particular, $\Delta$ includes correction terms that account for the difference between the approximations used in \eqs{assumptions}, and exact QFT, thus making \eq{488*} an exact expression for $K^{(2)}$. It is important that none of the of $2q2\bar q$-reducible diagrams in the last term of \eq{488*}  are overcounted, therefore $\Delta$ should not contain counter-terms for eliminating overcounting.\footnote{
 In fact our choice of the last term of \eq{488*} in this note is motivated by physics arguments and the possibility of close comparison with existing studies.}
 As such, $\Delta$ can be used in future studies to take into account effects such as one-gluon exchange, one-meson exchange, etc. 


\subsection{Tetraquark equations with coupling to $q\q$ states}

Equation  (\ref{id-q-tet}) constitutes the matrix form of the tetraquark equations without coupling to $q\q$ states. It expresses the column matrix $\phi$ of tetraquark form factors $\phi_M$ and $\phi_D$, in terms of potentials contained in matrix $V$. To derive the corresponding tetraquark equations {\em with} coupling to $q\q$ states, we simply use  the kernel $K^{(2)}$ of \eq{488*} in  \eq{2b-tetr}, the bound state equation for the tetraquark form factor $\Gamma^*$:
\begin{align}
\Gamma^*&=K^{(2)}G_0^{(2)}\Gamma^*    \nn
&=\left[\Delta +\bar NG_0^M\left(1-VG_0^M\right)^{-1}N\right] G_0^{(2)}\Gamma^*  \nn
&=\Delta G_0^{(2)}\Gamma^*  +\bar N G_0^M \Phi  \eqn{hel*}
\end{align}
where
\be
\Phi=V G_0^M \Phi +NG_0^{(2)}\Gamma^*.   \eqn{Phi}
\ee
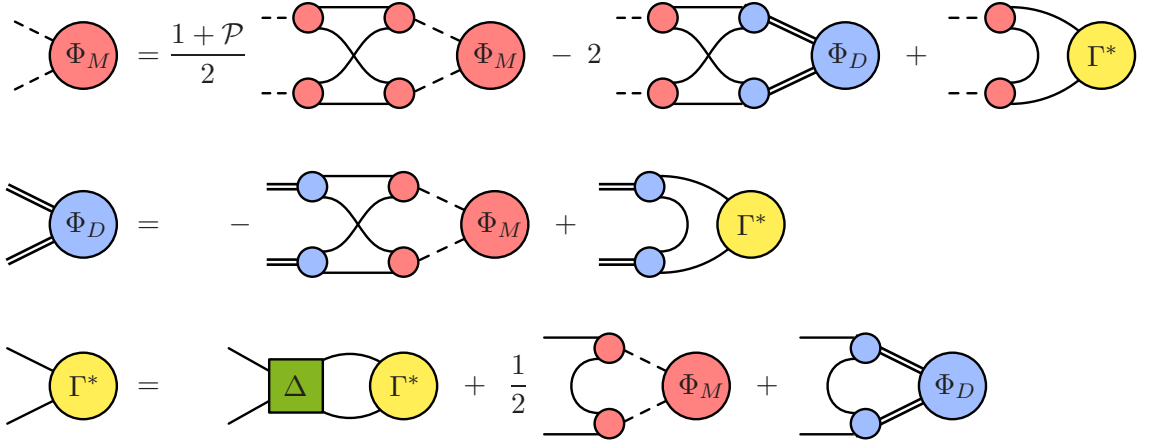
\begin{figure}[t]
\begin{center}
\begin{fmffile}{Phi}
\begin{align*}
\parbox{10mm}{
\begin{fmfgraph*}(10,15)
\fmfstraight
\fmfleftn{l}{7}\fmfrightn{r}{7}\fmfbottomn{b}{6}\fmftopn{t}{6}
\fmf{phantom}{l2,mb2,r2}
\fmf{phantom}{l6,mt2,r6}
\fmffreeze
\fmf{phantom}{r2,phi,r6}
\fmffreeze
\fmf{dashes}{phi,l2}
\fmf{dashes}{phi,l6}
\fmfv{d.sh=circle,d.f=empty,d.si=25,label=$\hspace{-4.5mm}\Phi_M$,background=(1,,.51,,.5)}{phi}
\end{fmfgraph*}}
\hspace{6mm} &=\frac{1+{\cal P}}{2}\hspace{2mm}
\parbox{30mm}{
\begin{fmfgraph*}(30,15)
\fmfstraight
\fmfleftn{l}{7}\fmfrightn{r}{7}\fmfbottomn{b}{6}\fmftopn{t}{6}
\fmf{phantom}{l2,vb1,mb1,vb2,mb2,r2}
\fmf{phantom}{l6,vt1,mt1,vt2,mt2,r6}
\fmf{phantom}{l2,xb1,mb1,xb2,mb2,r2}
\fmf{phantom}{l6,xt1,mt1,xt2,mt2,r6}
\fmf{phantom}{l2,yb1,mb1,yb2,mb2,r2}
\fmf{phantom}{l6,yt1,mt1,yt2,mt2,r6}
\fmffreeze
\fmf{phantom}{r2,phi,r6}
\fmffreeze
\fmfshift{4 right}{xb1}
\fmfshift{4 down}{xb1}
\fmfshift{4 left}{xt2}
\fmfshift{4 down}{xt2}
\fmfshift{4 right}{xt1}
\fmfshift{4 down}{xt1}
\fmfshift{4 left}{xb2}
\fmfshift{4 down}{xb2}
\fmfshift{4 right}{yb1}
\fmfshift{4 up}{yb1}
\fmfshift{4 left}{yt2}
\fmfshift{4 up}{yt2}
\fmfshift{4 right}{yt1}
\fmfshift{4 up}{yt1}
\fmfshift{4 left}{yb2}
\fmfshift{4 up}{yb2}
\fmfv{d.sh=circle,d.f=empty,d.si=11,background=(1,,.51,,.5)}{vb1}
\fmfv{d.sh=circle,d.f=empty,d.si=11,background=(1,,.51,,.5)}{vb2}
\fmfv{d.sh=circle,d.f=empty,d.si=11,background=(1,,.51,,.5)}{vt1}
\fmfv{d.sh=circle,d.f=empty,d.si=11,background=(1,,.51,,.5)}{vt2}
\fmf{dashes}{l2,vb1}
\fmf{dashes}{l6,vt1}
\fmf{dashes}{phi,xb2}
\fmf{dashes}{phi,yt2}
\fmf{phantom}{vb1,vb2}
\fmf{phantom}{vt1,vt2}
\fmfi{plain}{vloc(__xt2) {left}..tension 1 ..{left}vloc(__yb1)}
\fmfi{plain}{vloc(__xt1) {right}.. tension 1 ..{right}vloc(__yb2)}
\fmfi{plain}{vloc(__yt2) .. vloc(__yt1)}
\fmfi{plain}{vloc(__xb1) .. vloc(__xb2)}
\fmfv{d.sh=circle,d.f=empty,d.si=25,label=$\hspace{-4.5mm}\Phi_M$,background=(1,,.51,,.5)}{phi}
\end{fmfgraph*}}
\hspace{7mm}-\hspace{1mm}2\hspace{2mm}
\parbox{30mm}{
\begin{fmfgraph*}(30,15)
\fmfstraight
\fmfleftn{l}{7}\fmfrightn{r}{7}\fmfbottomn{b}{6}\fmftopn{t}{6}
\fmf{phantom}{l2,vb1,mb1,vb2,mb2,r2}
\fmf{phantom}{l6,vt1,mt1,vt2,mt2,r6}
\fmf{phantom}{l2,xb1,mb1,xb2,mb2,r2}
\fmf{phantom}{l6,xt1,mt1,xt2,mt2,r6}
\fmf{phantom}{l2,yb1,mb1,yb2,mb2,r2}
\fmf{phantom}{l6,yt1,mt1,yt2,mt2,r6}
\fmffreeze
\fmf{phantom}{r2,phi,r6}
\fmffreeze
\fmfshift{4 right}{xb1}
\fmfshift{4 down}{xb1}
\fmfshift{4 left}{xt2}
\fmfshift{4 down}{xt2}
\fmfshift{4 right}{xt1}
\fmfshift{4 down}{xt1}
\fmfshift{4 left}{xb2}
\fmfshift{4 down}{xb2}
\fmfshift{4 right}{yb1}
\fmfshift{4 up}{yb1}
\fmfshift{4 left}{yt2}
\fmfshift{4 up}{yt2}
\fmfshift{4 right}{yt1}
\fmfshift{4 up}{yt1}
\fmfshift{4 left}{yb2}
\fmfshift{4 up}{yb2}
\fmfv{d.sh=circle,d.f=empty,d.si=11,background=(1,,.51,,.5)}{vb1}
\fmfv{d.sh=circle,d.f=empty,d.si=11,background=(.6235,,.7412,,1)}{vb2}
\fmfv{d.sh=circle,d.f=empty,d.si=11,background=(1,,.51,,.5)}{vt1}
\fmfv{d.sh=circle,d.f=empty,d.si=11,background=(.6235,,.7412,,1)}{vt2}
\fmf{dashes}{l2,vb1}
\fmf{dashes}{l6,vt1}
\fmf{dbl_plain}{phi,xb2}
\fmf{dbl_plain}{phi,yt2}
\fmf{phantom}{vb1,vb2}
\fmf{phantom}{vt1,vt2}
\fmfi{plain}{vloc(__xt2) {left}..tension 1 ..{left}vloc(__yb1)}
\fmfi{plain}{vloc(__xt1) {right}.. tension 1 ..{right}vloc(__yb2)}
\fmfi{plain}{vloc(__yt2) .. vloc(__yt1)}
\fmfi{plain}{vloc(__xb1) .. vloc(__xb2)}
\fmfv{d.sh=circle,d.f=empty,d.si=25,label=$\hspace{-4.5mm}\Phi_D$,background=(.6235,,.7412,,1)}{phi}
\end{fmfgraph*}}
\hspace{7mm}+\hspace{2mm}
\parbox{20mm}{
\begin{fmfgraph*}(20,15)
\fmfstraight
\fmfleftn{l}{7}\fmfrightn{r}{7}\fmfbottomn{b}{4}\fmftopn{t}{4}
\fmf{phantom}{l2,vb1,mb1,r2}
\fmf{phantom}{l6,vt1,mt1,r6}
\fmf{phantom}{l2,xb1,mb1,r2}
\fmf{phantom}{l6,xt1,mt1,r6}
\fmf{phantom}{l2,yb1,mb1,r2}
\fmf{phantom}{l6,yt1,mt1,r6}
\fmffreeze
\fmf{phantom}{r2,phi,r6}
\fmffreeze
\fmfshift{4 right}{xb1}
\fmfshift{4 down}{xb1}
\fmfshift{4 right}{xt1}
\fmfshift{4 down}{xt1}
\fmfshift{4 right}{yb1}
\fmfshift{4 up}{yb1}
\fmfshift{4 right}{yt1}
\fmfshift{4 up}{yt1}
\fmfv{d.sh=circle,d.f=empty,d.si=11,background=(1,,.51,,.5)}{vb1}
\fmfv{d.sh=circle,d.f=empty,d.si=11,background=(1,,.51,,.5)}{vt1}
\fmf{dashes}{l2,vb1}
\fmf{dashes}{l6,vt1}
\fmfi{plain}{vloc(__yb1) {right}.. tension 1 ..{left}vloc(__xt1)}
\fmfi{plain}{vloc(__yt1) {right}.. vloc(__phi)}
\fmfi{plain}{vloc(__xb1){right} .. vloc(__phi)}
\fmfv{d.sh=circle,d.f=empty,d.si=25,label=$\hspace{-4.mm}\Gamma^*$,background=(1,,.9333,,.333)}{phi}
\end{fmfgraph*}}\\[6mm]
\parbox{10mm}{
\begin{fmfgraph*}(10,15)
\fmfstraight
\fmfleftn{l}{7}\fmfrightn{r}{7}\fmfbottomn{b}{6}\fmftopn{t}{6}
\fmf{phantom}{l2,mb2,r2}
\fmf{phantom}{l6,mt2,r6}
\fmffreeze
\fmf{phantom}{r2,phi,r6}
\fmffreeze
\fmf{dbl_plain}{phi,l2}
\fmf{dbl_plain}{phi,l6}
\fmfv{d.sh=circle,d.f=empty,d.si=25,label=$\hspace{-4.5mm}\Phi_D$,background=(.6235,,.7412,,1)}{phi}
\end{fmfgraph*}}
\hspace{6mm} &=\hspace{8mm}  - \hspace{2mm}
\parbox{30mm}{
\begin{fmfgraph*}(30,15)
\fmfstraight
\fmfleftn{l}{7}\fmfrightn{r}{7}\fmfbottomn{b}{6}\fmftopn{t}{6}
\fmf{phantom}{l2,vb1,mb1,vb2,mb2,r2}
\fmf{phantom}{l6,vt1,mt1,vt2,mt2,r6}
\fmf{phantom}{l2,xb1,mb1,xb2,mb2,r2}
\fmf{phantom}{l6,xt1,mt1,xt2,mt2,r6}
\fmf{phantom}{l2,yb1,mb1,yb2,mb2,r2}
\fmf{phantom}{l6,yt1,mt1,yt2,mt2,r6}
\fmffreeze
\fmf{phantom}{r2,phi,r6}
\fmffreeze
\fmfshift{4 right}{xb1}
\fmfshift{4 down}{xb1}
\fmfshift{4 left}{xt2}
\fmfshift{4 down}{xt2}
\fmfshift{4 right}{xt1}
\fmfshift{4 down}{xt1}
\fmfshift{4 left}{xb2}
\fmfshift{4 down}{xb2}
\fmfshift{4 right}{yb1}
\fmfshift{4 up}{yb1}
\fmfshift{4 left}{yt2}
\fmfshift{4 up}{yt2}
\fmfshift{4 right}{yt1}
\fmfshift{4 up}{yt1}
\fmfshift{4 left}{yb2}
\fmfshift{4 up}{yb2}
\fmfv{d.sh=circle,d.f=empty,d.si=11,background=(.6235,,.7412,,1)}{vb1}
\fmfv{d.sh=circle,d.f=empty,d.si=11,background=(1,,.51,,.5)}{vb2}
\fmfv{d.sh=circle,d.f=empty,d.si=11,background=(.6235,,.7412,,1)}{vt1}
\fmfv{d.sh=circle,d.f=empty,d.si=11,background=(1,,.51,,.5)}{vt2}
\fmf{dbl_plain}{l2,vb1}
\fmf{dbl_plain}{l6,vt1}
\fmf{dashes}{phi,xb2}
\fmf{dashes}{phi,yt2}
\fmf{phantom}{vb1,vb2}
\fmf{phantom}{vt1,vt2}
\fmfi{plain}{vloc(__xt2) {left}..tension 1 ..{left}vloc(__yb1)}
\fmfi{plain}{vloc(__xt1) {right}.. tension 1 ..{right}vloc(__yb2)}
\fmfi{plain}{vloc(__yt2) .. vloc(__yt1)}
\fmfi{plain}{vloc(__xb1) .. vloc(__xb2)}
\fmfv{d.sh=circle,d.f=empty,d.si=25,label=$\hspace{-4.5mm}\Phi_M$,background=(1,,.51,,.5)}{phi}
\end{fmfgraph*}}
\hspace{7mm}+\hspace{2mm}
\parbox{20mm}{
\begin{fmfgraph*}(20,15)
\fmfstraight
\fmfleftn{l}{7}\fmfrightn{r}{7}\fmfbottomn{b}{4}\fmftopn{t}{4}
\fmf{phantom}{l2,vb1,mb1,r2}
\fmf{phantom}{l6,vt1,mt1,r6}
\fmf{phantom}{l2,xb1,mb1,r2}
\fmf{phantom}{l6,xt1,mt1,r6}
\fmf{phantom}{l2,yb1,mb1,r2}
\fmf{phantom}{l6,yt1,mt1,r6}
\fmffreeze
\fmf{phantom}{r2,phi,r6}
\fmffreeze
\fmfshift{4 right}{xb1}
\fmfshift{4 down}{xb1}
\fmfshift{4 right}{xt1}
\fmfshift{4 down}{xt1}
\fmfshift{4 right}{yb1}
\fmfshift{4 up}{yb1}
\fmfshift{4 right}{yt1}
\fmfshift{4 up}{yt1}
\fmfv{d.sh=circle,d.f=empty,d.si=11,background=(.6235,,.7412,,1)}{vb1}
\fmfv{d.sh=circle,d.f=empty,d.si=11,background=(.6235,,.7412,,1)}{vt1}
\fmf{dbl_plain}{l2,vb1}
\fmf{dbl_plain}{l6,vt1}
\fmfi{plain}{vloc(__yb1) {right}.. tension 1 ..{left}vloc(__xt1)}
\fmfi{plain}{vloc(__yt1) {right}.. vloc(__phi)}
\fmfi{plain}{vloc(__xb1){right} .. vloc(__phi)}
\fmfv{d.sh=circle,d.f=empty,d.si=25,label=$\hspace{-4.mm}\Gamma^*$,background=(1,,.9333,,.333)}{phi}
\end{fmfgraph*}}\\[5mm]
\parbox{10mm}{
\begin{fmfgraph*}(10,15)
\fmfstraight
\fmfleftn{l}{7}\fmfrightn{r}{7}\fmfbottomn{b}{3}\fmftopn{t}{3}
\fmf{phantom}{l2,mb2,r2}
\fmf{phantom}{l6,mt2,r6}
\fmffreeze
\fmf{phantom}{r2,phi,r6}
\fmffreeze
\fmf{plain}{phi,l2}
\fmf{plain}{phi,l6}
\fmfv{d.sh=circle,d.f=empty,d.si=25,label=$\hspace{-4.mm}\Gamma^*$,background=(1,,.9333,,.333)}{phi}
\end{fmfgraph*}}
\hspace{6mm} &=\hspace{8mm}
\parbox{23mm}{
\begin{fmfgraph*}(23,15)
\fmfstraight
\fmfleftn{l}{7}\fmfrightn{r}{7}\fmfbottomn{b}{5}\fmftopn{t}{5}
\fmf{phantom}{l2,mb1,vb1,mb2,r2}
\fmf{phantom}{l6,mt1,vt1,mt2,r6}
\fmf{phantom,tension=.3}{l2,vb1}
\fmf{phantom,tension=.3}{l6,vt1}
\fmffreeze
\fmf{phantom}{r2,phi,r6}
\fmffreeze
\fmf{phantom}{vb1,delta,vt1}
\fmffreeze
\fmf{plain}{l2,delta}
\fmf{plain}{l6,delta}
\fmf{plain,left=.6}{delta,phi}
\fmf{plain,right=.6}{delta,phi}
\fmfv{d.sh=circle,d.f=empty,d.si=25,label=$\hspace{-4.mm}\Gamma^*$,background=(1,,.9333,,.333)}{phi}
\fmfv{d.sh=square,d.f=empty,d.si=20,background=(0.5216,,.7137,,.13333)}{delta}
\fmfiv{l=$\Delta$,l.a=180,l.d=.05w}{c}
\end{fmfgraph*}}
\hspace{7mm}+\hspace{2mm}\frac{1}{2}\hspace{2mm}
\parbox{20mm}{
\begin{fmfgraph*}(20,15)
\fmfstraight
\fmfleftn{l}{7}\fmfrightn{r}{7}\fmfbottomn{b}{5}\fmftopn{t}{5}
\fmf{phantom}{l2,mb1,vb2,mb2,r2}
\fmf{phantom}{l6,mt1,vt2,mt2,r6}
\fmf{phantom}{l2,mb1,xb2,mb2,r2}
\fmf{phantom}{l6,mt1,xt2,mt2,r6}
\fmf{phantom}{l2,mb1,yb2,mb2,r2}
\fmf{phantom}{l6,mt1,yt2,mt2,r6}
\fmf{phantom,tension=3}{l2,mb1}
\fmf{phantom,tension=3}{l6,mt1}
\fmffreeze
\fmf{phantom}{r2,phi,r6}
\fmffreeze
\fmfshift{4 down}{l2}
\fmfshift{4 up}{l6}
\fmfshift{4 left}{xt2}
\fmfshift{4 down}{xt2}
\fmfshift{4 left}{xb2}
\fmfshift{4 down}{xb2}
\fmfshift{4 left}{yt2}
\fmfshift{4 up}{yt2}
\fmfshift{4 left}{yb2}
\fmfshift{4 up}{yb2}
\fmfv{d.sh=circle,d.f=empty,d.si=11,background=(1,,.51,,.5)}{vb2}
\fmfv{d.sh=circle,d.f=empty,d.si=11,background=(1,,.51,,.5)}{vt2}
\fmf{dashes}{phi,xb2}
\fmf{dashes}{phi,yt2}
\fmfi{plain}{vloc(__xt2) {left}..tension 1 ..{right}vloc(__yb2)}
\fmfi{plain}{vloc(__yt2) .. vloc(__l6)}
\fmfi{plain}{vloc(__l2) .. vloc(__xb2)}
\fmfv{d.sh=circle,d.f=empty,d.si=25,label=$\hspace{-4.5mm}\Phi_M$,background=(1,,.51,,.5)}{phi}
\end{fmfgraph*}}
\hspace{7mm}+\hspace{2mm}
\parbox{20mm}{
\begin{fmfgraph*}(20,15)
\fmfstraight
\fmfleftn{l}{7}\fmfrightn{r}{7}\fmfbottomn{b}{5}\fmftopn{t}{5}
\fmf{phantom}{l2,mb1,vb2,mb2,r2}
\fmf{phantom}{l6,mt1,vt2,mt2,r6}
\fmf{phantom}{l2,mb1,xb2,mb2,r2}
\fmf{phantom}{l6,mt1,xt2,mt2,r6}
\fmf{phantom}{l2,mb1,yb2,mb2,r2}
\fmf{phantom}{l6,mt1,yt2,mt2,r6}
\fmf{phantom,tension=3}{l2,mb1}
\fmf{phantom,tension=3}{l6,mt1}
\fmffreeze
\fmf{phantom}{r2,phi,r6}
\fmffreeze
\fmfshift{4 down}{l2}
\fmfshift{4 up}{l6}
\fmfshift{4 left}{xt2}
\fmfshift{4 down}{xt2}
\fmfshift{4 left}{xb2}
\fmfshift{4 down}{xb2}
\fmfshift{4 left}{yt2}
\fmfshift{4 up}{yt2}
\fmfshift{4 left}{yb2}
\fmfshift{4 up}{yb2}
\fmfv{d.sh=circle,d.f=empty,d.si=11,background=(.6235,,.7412,,1)}{vb2}
\fmfv{d.sh=circle,d.f=empty,d.si=11,background=(.6235,,.7412,,1)}{vt2}
\fmf{dbl_plain}{phi,xb2}
\fmf{dbl_plain}{phi,yt2}
\fmfi{plain}{vloc(__xt2) {left}..tension 1 ..{right}vloc(__yb2)}
\fmfi{plain}{vloc(__yt2) .. vloc(__l6)}
\fmfi{plain}{vloc(__l2) .. vloc(__xb2)}
\fmfv{d.sh=circle,d.f=empty,d.si=25,label=$\hspace{-4.5mm}\Phi_D$,background=(.6235,,.7412,,1)}{phi}
\end{fmfgraph*}}
\end{align*}
\end{fmffile}   
\vspace{3mm}

\caption{\fign{eqns}  Illustration of the tetraquark equations, \eqs{4eq-DD***}, with coupling to $q\q$ states included. Tetraquark form factors $\Phi_M$ (displayed in red) couple to two mesons (dashed lines), tetraquark form factors $\Phi_D$ (displayed in blue) couple to diquark and antidiquark states (double-lines), and the tetraquark form factors $\Gamma^*$ (displayed in yellow) couple to $q\q$ states (solid lines). The amplitude $\Delta$ (displayed in green) represents all contributions to the $q\q$ kernel $K^{(2)}$ that are not included in the last term of \eq{488*}.}
\end{center}
\end{figure}
Equation (\ref{Phi}) is the matrix form of the sought-after tetraquark equations with coupling to $q\q$ states.  It expresses the column matrix $\Phi$ of tetraquark form factors $\Phi_M$ and $\Phi_D$ in terms of both the potentials contained in matrix $V$, and the tetraquark form factor $\Gamma^*$ describing the disintegration of a tetraquark into a $q\q$ pair. We can write \eq{Phi} explicitly as
\begin{align}
\begin{pmatrix} \Phi_M \\ \Phi_D \end{pmatrix} 
&=
\begin{pmatrix} (1+{\cal P})\bar\Gamma_M\Gf_0{\cal P}_{34}\Gamma_M& -2\bar\Gamma_M \Gf_0 \Gamma_D \\
-2\bar\Gamma_D\Gf_0\Gamma_M & 0 \end{pmatrix} 
\begin{pmatrix} \frac{1}{2}MM & 0\\  0 & D\bar D \end{pmatrix} 
\begin{pmatrix} \Phi_M \\ \Phi_D \end{pmatrix} 
+ \begin{pmatrix} N_M \\ N_D \end{pmatrix} G_0^{(2)}\Gamma^*  \eqn{eqns-mat}
\end{align}
where $\Gamma_M=\Gamma_{13} \Gamma_{24}$,  $\bar\Gamma_M =\bar\Gamma_{13} \bar\Gamma_{24}$, 
 $\Gamma_D=  \Gamma_{12} \Gamma_{34}$,    
 $\bar\Gamma_D = \bar\Gamma_{12} \bar\Gamma_{34} $, and ${\cal P}_{ij}$ is the operator exchanging quarks $i$ and $j$, therefore
\begin{align}
\bar\Gamma_MG_0^{(4)}{\cal P}_{34}\Gamma_M =\bar\Gamma_{13} \bar\Gamma_{24}G_0^{(4)}\Gamma_{14} \Gamma_{23},\hspace{1cm}
\bar\Gamma_MG_0^{(4)}\Gamma_D= \bar\Gamma_{13} \bar\Gamma_{24}G_0^{(4)}\Gamma_{12} \Gamma_{34}.   \eqn{fin-notat}
\end{align}
Thus the tetraquark equations with coupling to $q\q$ included take the form of three coupled equations
\begin{subequations}  \eqn{4eq-DD***}
\begin{align}
\Phi_M&=
(1+{\cal P})\bar\Gamma_M G^{(4)}_0{\cal P}_{34}\Gamma_M \frac{MM}{2}\Phi_M  
-2\bar\Gamma_M G^{(4)}_0\Gamma_D D\bar D\Phi_D + N_MG_0^{(2)}\Gamma^*,
 \\[2mm]
\Phi_D&=-2\bar\Gamma_D G^{(4)}_0
\Gamma_M\frac{MM}{2}\Phi_M  +N_DG_0^{(2)}\Gamma^*  ,
 \\[2mm]
\Gamma^*&=\Delta G_0^{(2)}\Gamma^*+ \bar N_M\frac{MM}{2} \Phi_M +\bar N_D D\bar D \Phi_D ,   \eqn{c}
\end{align}
\end{subequations}
which are illustrated in \fig{eqns}. 
Since $\Delta$ is defined in a way that makes the expression used for $K^{(2)}$ exact,  \eqs{4eq-DD***} represent the most general form of the tetraquark equations in QFT.

\section{Conclusions}

We have derived a set of covariant coupled equations for the tetraquark, \eqs{4eq-DD***}, using a model where the two-body $q \q$, $q q$, and $\q\q$ interactions are dominated by the formation of a meson, a diquark, and an antidiquark, respectively. Nevertheless, \eqs{4eq-DD***} constitute the most general form of the tetraquark equations in QFT since all differences between the model used and exact QFT are accounted for by correction terms formally included in the term $\Delta$.  These equations determine the form factors $\Phi_M$, $\Phi_D$, and $\Gamma^*$ of the tetraquark, describing its disintegration into two identical mesons, a diquark-antidiquark pair, and a quark-antiquark pair. As such, they extend the purely four-body ($4q$) tetraquark equations of Ref.\ \cite{Heupel:2012ua} to include coupling to two-body($2q$)  $q\q$ states.

The motivation for the present work comes from the need to have exact quantum field theoretic equations describing the tetraquark, but formulated for the case where the dynamics is dominated by meson-meson and diquark-antidiquark components. This is especially important in view of the lack of agreement between two previous attempts to calculate tetraquark equations with $4q$-$2q$ mixing. The first of these was our derivation of 2014 \cite{Kvinikhidze:2014yqa} using a careful but involved incorporation of disconnected $q\q$ interactions as a means of incorporating $q\q$ annihilation into a $4q$ description. The second of these was a recent derivation \cite{Santowsky:2020pwd} where coupling to $2q$ channels was included phenomenologically, and where some doubt was expressed regarding the incorporation of disconnected $q\q$ interactions. Our present derivation of \eqs{4eq-DD***} has therefore been based on a method that avoids any explicit introduction of disconnected $q\q$ interactions, and which, in the absence of approximations for $\Delta$,  provides an exact field-theoretic description. It is therefore gratifying to note that in the absence of the term $\Delta$, \eqs{4eq-DD***} coincide with the equations derived by us in Ref.\  \cite{Kvinikhidze:2014yqa}. Indeed, settting $\Delta=0$ in \eq{c} and substituting into \eq{eqns-mat} gives $\Phi$ in the form presented in Ref.\ \cite{Kvinikhidze:2014yqa}:\footnote{The expression in the square bracket in \eq{blank} may appear to come with an opposite sign in Ref.\ \cite{Kvinikhidze:2014yqa}, but this is not the case as the definitions of $\bar \Gamma_M$, $\Gamma_M$, $\bar \Gamma_D$, $\Gamma_D$, $\bar N$, and $N$ used in Ref.\ \cite{Kvinikhidze:2014yqa} differ from the ones used here.}

\begin{align}
\Phi&=\left[\left( \begin{array}{cc} (1+{\cal P}) \bar\Gamma_MG_0^{(4)}{\cal P}_{34}\Gamma_M
& -2\bar\Gamma_MG_0^{(4)}\Gamma_D\\
-2 \bar\Gamma_DG_0^{(4)}\Gamma_M &0 \end{array} \right)+NG_0^{(2)}\bar N
\right] 
 \left( \begin{array}{cc}\frac{1}{2}MM & 0 \\
 0 &D\bar D \end{array} \right)\Phi.  \eqn{blank}
\end{align}
Here $NG_0^{(2)}\bar N$ is the $q\bar q$ reducible part of the kernel which is denoted by $V_{q\bar q}$ in Ref.\ \cite{Kvinikhidze:2014yqa}. It accounts for the $q\bar q$ admixture through the $q\bar q$ propagator $G_0^{(2)}$.  By contrast, the tetraquark equations of Ref.\ \cite{Santowsky:2020pwd} are not consistent with the general form prescribed by \eqs{4eq-DD***}.

Finally, it is worth noting that in comparison with our previous derivation \cite{Kvinikhidze:2014yqa} , the approach taken in the present work allows for the derivation of the tetraquark equations in a much simpler and more clear way.

\begin{acknowledgments}

 A.N.K. was supported by the Shota Rustaveli National Science Foundation (Grant No. FR17-354).

\end{acknowledgments}
\newpage

\bibliography{/Users/phbb/Physics/Papers/refn} 

\end{document}